\documentclass[pagesize,12pt,a4paper]{scrartcl}
\usepackage{german}
\usepackage{amsmath}
\usepackage{amsthm}
\usepackage{amssymb}
\usepackage{amsxtra}
\usepackage[squaren]{SIunits}
\usepackage{graphicx}
\usepackage{braket}
\usepackage{capt-of}
\usepackage{ae}
\usepackage{wasysym}
\usepackage{makeidx}
\usepackage{inputenc}
\usepackage[T1]{fontenc}
\usepackage{url}
\usepackage{eepic}
\usepackage{mathrsfs}
\usepackage{comment}
\usepackage{mparhack}
\newcommand{\ir}{\mathrm{i}}
\newcommand{\e}{\mathrm{e}}

\renewcommand{\jmath}{j}
\newcommand{\longpage}{\enlargethispage{1\baselineskip}}

\DeclareMathOperator{\dv}{d}

\DeclareMathOperator*{\diag}{diag}
\ifx\KOMAScript\undefined%
  \DeclareRobustCommand{\KOMAScript}{\textsf{K\kern.05em O\kern.05em%
      M\kern.05em A\kern.1em-\kern.1em Script}}
\fi
\newlength{\help}
\newlength{\minuslaenge}
\newtheoremstyle{note}
  {3pt}
  {3pt}
  {\rmshape}
  {}
  {\bfseries}
  {:}
  {.5em}
  {}

\theoremstyle{note}

\makeatletter

 \def\vec#1{\ensuremath{\mathchoice
                     {\mbox{\boldmath$\displaystyle\mathbf{#1}$}}
                     {\mbox{\boldmath$\textstyle\mathbf{#1}$}}
                     {\mbox{\boldmath$\scriptstyle\mathbf{#1}$}}
                     {\mbox{\boldmath$\scriptscriptstyle\mathbf{#1}$}}}}%
\makeatother

\begin{document}

  \title{ 
          Heisenberg versus the Covariant String 
}
  \author{Norbert Dragon and Florian Oppermann\\
          Institut f\"ur Theoretische Physik\\
          Leibniz Universit\"at Hannover 
}
\date{}

\maketitle

\begin{abstract} 
A Poincar\'e multiplet of mass eigenstates $\bigl(P^2 - m^2\bigr)\Psi = 0$ cannot be a
subspace of a space
with a $D$-vector position operator $X=(X_0,\dots X_{D-1})$:
the Heisenberg algebra $[P^m, X_n] = \ir \delta^m{}_n$
implies by a simple argument that each Poincar\'e multiplet of definite mass vanishes. 

The same conclusion follows from the Stone-von Neumann theorem.

In a quantum theory the constraint of an absolutely continuous spectrum to a lower dimensional submanifold yields zero even if Dirac's treatment
of the corresponding classical constraint defines a symplectic submanifold with a consistent corresponding quantum model. 
Its Hilbert space is not a subspace of the unconstrained theory. Hence the operator relations of the unconstrained model
need not carry over to the constrained model.

Our argument excludes quantized worldline models of relativistic particles and  the physical states of the covariant quantum string.

We correct misconceptions about the generators of Lorentz transformations acting on particles.

\end{abstract}

\newpage

\section{Introduction}
The momentum $P=(P^0,\dots P^{D-1})$ generates the unitary 
representation $U_a = \e^{\ir\, P \,a}$
of  translations in space\-time. This makes models tempting which contain in addition a spacetime position operator \mbox{$X=(X_0,\dots X_{D-1})$}, 
which Lorentz transforms as a $D$-vector and which is translated,
\begin{equation} 
\label{xtrans}
\e^{\ir\, a\, P} X_n \,\e^{-\ir\, a\,  P}= X_n - a_n\ ,\   a \in \mathbb R^D\ . 
\end{equation}
Functions of $X$ such as $V_b(X ) = \e^{\ir\,b\, X}= \colon V_b$ are shifted,
$U_a \e^{\ir\,b\, X} U_a{}^{-1}=\e^{\ir\,b\, (X-a)} $. 
These are the Weyl relations 
\begin{equation}
\label{weyl}
U_a V_b =V_b U_a\, \e^{-\ir\, a\, b}\ ,\ U_a U_b =U_{a+b}\ ,\ V_a V_b =V_{a+b}\ . 
\end{equation}
Their generators satisfy the Heisenberg Lie algebra
\begin{equation}
\label{heisenberg}
 [P^m, P^n]=0 = [X_m, X_n]\ , \  [ P^n,X_m]= \ir\,\delta^n{}_m\ .
\end{equation}

Differentiating $V_b U_a V_b{}^{-1}=U_a\,\e^{\ir a b}$ with respect to $a_m$ at $a=0$ shows
\begin{equation}
\label{momentcont}
\e^{\ir\,b\, X} P^m \e^{-\ir\,b\, X} =P^m + b^m\ ,\  b \in \mathbb R^D\ . 
\end{equation}

Thus by (\ref{xtrans}) and (\ref{momentcont})
the operators $X_n$ and $P^m$ are unitarily equivalent to the shifted operators. 
So their spectra are invariant under shifts and consist of the continuum $\mathbb R^D$.

However, as observed invariably the possible momenta of relativistic particles do not fill a $D$-dimensional continuum but
are restricted to mass shells. Their states are spanned by multiplets (spaces with an irreducible unitary repre\-sentation of the Poincar\'e group with a definite scalar product) 
with discrete masses.
Though this discrepancy in string theories 
was observed \cite{bahns, dimock, grundling} it was not considered a severe fault. Lecture notes and books e.g. \cite{arutyunov, witten,thooft} 
claimed that multiplets of definite masses span a subspace, the space of physical states, of a larger space with a
unitary representation of the Weyl relations.

We show:  This is untenable. The spacetime Heisenberg Lie algebra (\ref{heisenberg}) excludes any subspace with a definite mass $m$ whether $m$ vanishes or not.

To constrain in world line models the continuous momentum spectrum to mass shells yields zero, 
because the volume measure of a lower dimensional submanifold vanishes. This is very much different from constraints which select from discrete possibilities.

In particular by this reason of vanishing measure the Stone-von Neumann theorem excludes mass shells in the continuous momentum spectrum.

If theories, different from the worldline models or the covariant quantum string, contain only the \emph{spatial } part of the Heisenberg Lie algebra then this is
consistent with massive particles, $m > 0$. This is compatible with Lorentz covariance, even though of an unusual kind. Covariance does not require $\vec X$ be 
the spatial components of a $D$-vector.

Because Lorentz generators, which are constructed using (\ref{heisenberg}), do \emph{not} act on particles with a definite mass,
we specify the generators which do. 

\section{Absence of Mass Multiplets}

\textbf{Lemma:} \textit{A Poincar\'e multiplet of states $\Psi$ of a definite mass~$m$
\begin{equation}
\label{mass}
(P^2 - m^2)\Psi = 0\ ,\ P^2 = (P^0)^2 - \sum_{i=1}^{D-1}(P^i)^2\ ,
\end{equation}
cannot be a subspace of a space in which Heisenberg pairs $(i,j \in \set{1,\dots D-1})$ satisfy
\begin{equation} 
\label{heisenspace}
[P^i, P^j]=0 = [X^i, X^j]\ , \ [ P^i,X^j]= -\ir\,\delta^{ij}\ ,
\end{equation}
and commute with $P^0$
\begin{equation}
\label{innocent}
 [P^0, X^i] \stackrel{?}{=} 0 \ .
\end{equation}
}

{Proof:}
The space is the orthogonal sum of the mass multiplet and its complement.
All arbitrarily chosen states $\Psi$ and $\Phi$ of the multiplet are orthogonal to the complement and 
have a vanishing matrix element of the commutator $[(P^2 - m^2),X^1]= 2\ir P^1 $
\begin{equation}
\begin{aligned}
\braket{\Phi\, |\, [(P^2 - m^2),X^1]\, \Psi} & = 
\braket{\,(P^2 - m^2)\Phi \,|\, X^1 \Psi} - \braket{\Phi \,| \,X^1\, (P^2 - m^2) \Psi}\\
& = 0 - 0 =  2 \ir \braket{\Phi\, |\, P^1 \Psi}
\end{aligned}
\end{equation}

All scalar products of $P^1 \Psi$ vanish.
But the scalar product is nondegenerate, hence
\begin{equation}
P^1 \Psi = 0\ .
\end{equation}
Exchanging $\Phi$ and $\Psi$ in the argument, one also has $P^1 \Phi = 0$. As $P^1$ is hermitian
their matrix element of the commutator $[P^1,X^1]= - \ir$ vanishes
\begin{equation}
\braket{\Phi \,|\,[P^1,X^1]\, \Psi} = \braket{P^1 \Phi\, | \,X^1 \Psi} - \braket{\Phi \,| \,X^1 P^1 \Psi}
= 0 - 0 =  -\ir \braket{\Phi|\Psi}\ .
\end{equation}

All scalar products of $\Psi$ vanish, thus 
\begin{equation}
\label{argument}
\Psi=0\ .
\end{equation}
The state $\Psi$ was arbitrarily chosen from the mass multiplet, so there is none. 
\qed

The argument needs only the nondegeneracy of the scalar product, not its positivity, and does not require $X^1$ be hermitian.
It needs no assumptions about wavefunctions, which constitute the multiplet, nor their explicit scalar product.

Our lemma excludes quantized models  \cite{arutyunov,hanson,scherk} of free relativistic particles which classically traverse worldlines
$t\mapsto x(t)$ with action given by their length. Canonical quantization 
yields the Heisenberg algebra. The mass shell condition $(P^2-m^2)\Psi=0$  arises as con\-straint because of the re\-para\-me\-tri\-za\-tion invariance $t\mapsto t'(t)$.
But, no matter how suggestive, aesthetical and geometrical a classical system may be, this does not gua\-ran\-tee that canonical quantization yields a quantum model 
in which the constraint has a non\-va\-ni\-shing solution.

These models contain the algebra (\ref{heisenberg}) and declare to contain a  multiplet 
of definite mass. They lay claim to the name \lq relativistic particle\rq\ but contain none. 
This justifies to rename them \lq worldline models\rq\  to avoid the confusing statement that relativistic
particles do not exist. Experimentally, relativistic particles are verified beyond doubt, but the worldline models
fail to describe them whatever their denomination pretends.
Our lemma applies also to the covariant quantum string which postulates~(\ref{heisenberg}).\footnote{We reserve the name 'covariant string' to string
models with the algebra (\ref{heisenberg}). This article does not deal with the light cone string which employs only a subalgebra.} Its Hilbert space is claimed to be an orthogonal
sum of mass multiplets -- the physical states~-- and a complement. 
Our lemma excludes any multiplet of definite mass. The covariant quantum string has no physical states.\footnote{The
result is unchanged by identifying states which differ by spurious states.}

In particular, one cannot constrain the absolutely continuous spectrum of a quantum model to a 
lower dimensional submanifold:  each integral of a projection valued measure on a set of vanishing measure
yields zero. This is what vanishing projection measure means.

In a quantum model the solutions to a constraint of a continuous spectrum  vanish
even in case that Dirac's formulation \cite{dirac} of the corresponding classical constraint defines a symplectic submanifold
with a consistent corresponding quantum model. Its Hilbert space is not a subspace of the unconstrained model. In the constrained model
the operator relations of the unconstrained model need not hold.

\section{The Stone-von Neumann Theorem}

By the Stone-von Neumann theorem \cite[Theorem XI.84]{reed3}\footnote{There for $D=1$. The result carries over to finite $D$ \cite[Notes 8.10]{schmuedgen1}.}
each unitary representation of the Weyl relations (\ref{weyl}) is unitarily equivalent to the one in a Hilbert space~$\mathfrak L^2(\mathbb R^D)\times \mathcal N$
of states $\Psi: p \mapsto \Psi(p)$ which map $p\in \mathbb R^D$ almost everywhere to $\Psi(p)$ in some Hilbert space~$\mathcal N$. The unitary representation acts multiplicatively and by translation
\begin{equation}
\label{heisenrep}
\begin{gathered}
(U_a \Psi)(p)= \e^{\ir\, a\, p}\,\Psi(p)\,,\ (V_b \Psi)(p)= \Psi(p+b)\ ,\\ 
\braket{\Phi | \Psi}= \int\! \dv^D\!\!p \braket{\Phi(p)|\Psi(p)}_{\mathcal N}\ .
\end{gathered}
\end{equation}
By the theorem one is not free to choose a different scalar product which integrates not over $\mathbb R^D$ but only over a mass shell
with measure $\dv^{D-1}p / \sqrt{m^2 + \vec p^2}$.


The scalar product of $\mathfrak L^2(\mathbb R^D)$ implies that each multiplet of definite mass vanishes: 
$\mathfrak L^2(\mathbb R^D)$ is the space of equivalence classes of square integrable wave functions,
which are equivalent if the $D$-dimensional measure of the support of their difference vanishes,
\begin{equation}
\Psi= 0 \Leftrightarrow \forall \Phi: \int\! \dv^D\!\!p\, \braket{\Phi(p)|\Psi(p)}_{\mathcal N} = 0\ .
\end{equation}
All wave functions $\Psi_{\text{phys}}\in \mathfrak L^2(\mathbb R^D)$ with definite mass $m$ 
\begin{equation}
\label{restriction}
\Psi_{\text{phys}}(p)=
0\  \text{ if }\ p^0 \ne \sqrt{m^2+\vec p^2}
\end{equation}
only have a $(D-1)$-dimensional support of vanishing $D$-dimensional measure.  
They are equivalent to~$0$ and vanish.

Physical states are not elements of $\mathfrak L^2(\mathbb R^{D-1})$ obtained from $\mathfrak L^2(\mathbb R^{D})$  
by restriction to the mass shell:
restriction of equivalence classes is a linear map, it vanishes if applied to~$0.$  On $\mathfrak L^2(\mathbb R^{D})$ 
restriction to a mass shell vanishes altogether.
To realize the algebra in a space with a different measure is excluded by the Stone-von Neumann theorem.

If it needed another argument: the Heisenberg algebra (\ref{heisenberg}) is represented by the
hermitian operators
\begin{equation}
P^m \Psi(p) = p^m \Psi\ ,\ X_n \Psi(p) = -\ir\, \partial_{p^n}\Psi(p)\ .
\end{equation}
They generate an algebra which is defined on and maps to itself the Schwartz space $\mathcal S(\mathbb R^D, \mathcal N)$ of 
smooth functions $\Psi:\mathbb R^D \rightarrow \mathcal N$ which together with each of their derivatives decrease rapidly \cite{schmuedgen1}.
The only smooth function of $\mathbb R^D$ which vanishes outside mass shells is $\Psi = 0$.

\section{Consistent Spatial Position Operator}

\longpage

The disastrous, innocent looking relation $ [X^i,P^0] \stackrel{?}{=} 0$ (\ref{innocent}) in worldline models 
is incompatible with the Schr\" odinger equation $\ir\, \partial_t \Psi(t) = P^0 \Psi(t)$ 
for the motion of a massive relativistic particle. 
For its expected position $x^i(t)=\braket{\Psi(t) | X^i \Psi(t) }$ 
to change in the course of the time $t$
by the expected velocity $\partial_t x^i = v^i = \braket{\Psi|\bigl(P^i/P^0\bigr) \Psi}$
one has to have not (\ref{innocent}) but
\begin{equation}
\label{spaceenergy}
[X^i,P^0]=\ir \frac{P^i}{P^0} \Leftrightarrow [X^i, P^2 - m^2] = 0\ .
\end{equation}
It is this value 
which the commutator of~$X^i$ with $P^0$ must have in order to commute with the mass shell condition.
Moreover, (\ref{spaceenergy}) is required to justify the denomination \lq position operator\rq.
It entails the idea  that in the course of time the position of a particle changes with its velocity.

Poincar\'e covariance does not require the position operator~$\vec X$ be the spatial part of a $D$-vector:
$\vec X=\vec X_{\underline u} $ is the position operator used by an observer at rest with four-velocity $\underline u = (1,0,\dots)$.
Under spacetime translations $a$ and rotations~$R$ it transforms linear inhomogeneously\footnote{In an orthonormal basis our 
metric  is $\eta = \diag (1,-1,\dots, -1)$.}   
\begin{equation}
\e^{\ir a  P}\, \vec X \,\e^{-\ir a P} = \vec X + \vec a + a^0 \frac{\vec P}{P^0}\ ,\ 
U_{R} \vec X U_{R}{}^{-1}= R^{-1} \vec X\ ,
\end{equation}
where $U_{R}$ represents the rotation $R$ in Hilbert space.
 Observers boosted by $L_u$ to four-velocity~$u$ measure position with 
\begin{equation}
\label{xu}
\vec X_u = U_{L_u}\vec X_{\underline u} U_{L_u}{}^{-1}\ .
\end{equation}
 Under Lorentz transformations $\Lambda$
the position operators transform by Wigner rotation
\begin{equation}
\label{xuw}
U_\Lambda \vec X_u U_\Lambda{}^{-1}= W^{-1}(\Lambda,u) \vec X_{\Lambda u}\ , \ W(\Lambda, u) =  L_{\Lambda u}{}^{-1} \Lambda L_u\in \text{SO}(D-1)\ ,     
\end{equation}
in a Poincar\'e covariant way, even though $\vec X$ is not the spatial part of a $D$-vector. It is an element of a $(D-1)$-parameter set of
$(D-1)$-vectors.

\section{Lorentz Generators of Particles}

In terms of the algebra (\ref{heisenberg}) one can easily specify  generators $M_\omega = \omega^{mn} M_{mn}/2$ 
of Lorentz transformations $U_{\e^\omega} = \e^{-\ir M_{\omega}}$ \cite{thooft}, 
\begin{equation}
\label{lorentzcov}
-\ir M^{mn}\Psi \stackrel{?}{=} \ir (P^m X^n - P^n X^m)\Psi + \Gamma^{mn} \Psi\ , 
\end{equation}
where $\Gamma^{mn}$ are skew hermitian matrices which commute with $X$ and $P$ and represent the Lorentz Lie algebra
\begin{equation}
 [\Gamma^{mn}, \Gamma^{rs}] = -\eta^{mr}\Gamma^{ns}+\eta^{ms}\Gamma^{nr}+\eta^{nr}\Gamma^{ms}-\eta^{ns}\Gamma^{mr}
\end{equation}
as do the operators $l^{mn}= \ir (P^m X^n - P^n X^m)$.

Nonvanishing $\Gamma^{mn}$ can occur only in case the scalar product in spin space is indefinite, otherwise there are no finite dimensional, skew hermitian
matrices which generate the Lorentz group. 

However as our lemma shows, the operators $P$ and $X$ act not on particles with a definite mass
but in a space in which $P$ has the continuous spectrum $\mathbb R^D$ (\ref{momentcont}) with unbounded and also negative energies.
This is not the space of relativistic particles.



In the correct description, rather, massive one-particle particle states are momentum wave functions~$\Psi: \mathcal M_m \rightarrow \mathbb C^{d} $
 which map the massive shell, 
\begin{equation}
\mathcal M_m =\set{p:\,p^0 = \sqrt{m^2+\vec p^2}\,,\  \vec p \in \mathbb R^{D-1}} \subset \mathbb R^D\ , \  m > 0\ ,
\end{equation}
to some space $\mathbb C^{d}$, in which skew hermitian matrices $\Gamma_{ij}=-\Gamma_{ji}$ generate a $d$-dimensional unitary representation of SO$(D-1)$,
$(i,j,k,l\in \set{1,\dots ,D-1})$,
\begin{equation}
\label{sod-1}
 [\Gamma_{ij}, \Gamma_{kl}] = \delta_{ik}\Gamma_{jl}-\delta_{jk}\Gamma_{il}-\delta_{il}\Gamma_{jk}+\delta_{jl}\Gamma_{ik}\ .
\end{equation}
The generators of the Poincar\'e group \cite{mackey}  and the position operator map by\footnote{We use matrix notation 
and suppress indices of the components of $\Psi$ and~$\Gamma_{ij}$. Checking the Lorentz algebra observe $\sum_{i=1}^{D-1} p^ip^i=(p^0)^2 - m^2=(p^0+m)(p^0-m)$.}  
\begin{equation}
\begin{aligned}
\label{massiv}
(-\ir\, P^n \Psi)(p) & = -\ir\, p^n \Psi(p)\ ,\\
\bigl(-\ir M_{ij}\Psi\bigr)(p) & = -\bigl(p^i \partial_{p^j} - p^j \partial_{p^i}\bigr)\Psi(p) + \Gamma_{ij}\Psi(p)\ ,\\
\bigl(-\ir M_{0i}\Psi\bigr)(p) & = p^0\,\partial_{p^i}\Psi(p) + \Gamma_{ij}\frac{p^j}{p^0 + m}\Psi(p)\ ,\\
\bigl(-\ir X^i \Psi\bigr)(p) &= \partial_{p^i}\Psi(p) + \frac{p^i}{2(p^0)^2}\Psi(p)\ ,
\end{aligned}
\end{equation}
the Schwartz space $\mathcal S(\mathcal M_m, \mathbb C^d)$ of smooth states of rapid decrease to itself.
The generators are skew hermitian with respect to the Lorentz invariant measure $(\dv^{D-1}\!p)/p^0$. 

\longpage

They are equivariant: observers, Lorentz boosted by $L_u$, use the generators (\ref{xu}, \ref{xuw}) and
\begin{equation}
U_{L_u} M_{mn} U_{L_u}{}^{-1}=\bigl(L_u\bigr)^r{}_m \bigl(L_u\bigr)^s{}_n M_{rs}\ .
\end{equation}

The massless case is \emph{not} obtained by simply specifying  $m=0$ in the energy
$p^0=\sqrt{\vec p^2}$. 
Its inverse $1/p^0$ and therefore the invariant measure $(\dv^{D-1}\!p)/p^0$, the 
generators $M_{0i}$ and $X^i$ are singular at $\vec p=0$. There the energy is only continuous, not smooth.
The distinguished momentum $p=0$ is a fixed point of Lorentz transformations and \emph{not} invariant under translations. 
So there cannot exist generators $X^i$ of such translations: 
massless states do not allow the spatial Heisenberg algebra (\ref{heisenspace}). 

\longpage

The Lorentz generators of massless states turn out not to act on smooth functions of $\mathbb R^{D-1}$
but on smooth sections of a vector bundle over
$S^{D-2}\times \mathbb R$ which carries a representation of SO$(D-2)$ with generating matrices $\Gamma_{ij}$. 

In the coordinate patch $\mathcal U_N = \set{p: p^0=\sqrt{\vec p^2}, |\vec p| + p_z > 0}$
the sections are smooth functions $\Psi_N$ which the generators map 
 to ($p_z := p^{D-1}$,  $i,j,k\in \set{1,\dots D-2}$)\footnote{Checking 
the Lorentz algebra observe $\sum_{i=1}^{D-2} p^ip^i=|\vec p|^2 - (p_z)^2=(|\vec p| + p_z )(|\vec p| - p_z )$.}  
\begin{equation} 
\begin{aligned}
\bigl(-\ir M_{ij}\Psi\bigr)_N(p)&= - \bigl(p^i\partial_{p^j} - p^j\partial_{p^i}\bigr)\Psi_N(p) + \Gamma_{ij}\,\Psi_N(p)\ ,\\
\bigl(-\ir M_{zi}\Psi\bigr)_N(p)&= - \bigl(p_z\partial_{p^i} - p^i\partial_{p_z}\bigr)\Psi_N(p) + \Gamma_{ik}\frac{p^k}{|\vec p|+p_z}\,\Psi_N(p)\ ,\\
\bigl(-\ir M_{0i}\Psi\bigr)_N(p)&=  |\vec p|\partial_{p^i}\Psi_N(p) + \Gamma_{ik}\frac{p^k}{|\vec p|+p_z}\,\Psi_N(p)\ ,\\
\bigl(-\ir M_{0z}\Psi\bigr)_N(p)&=  |\vec p|\partial_{p_z}\Psi_N(p)\ . 
\end{aligned}
\end{equation}

The detailed discussion of massless particles, e.g. the relation of $\Psi_N$ to $\Psi_S$ which is smooth in 
$\mathcal U_S = \set{p: p^0=\sqrt{\vec p^2}, |\vec p| - p_z > 0}$ will be given elsewhere \cite{dragon}. 
Here it is only important that in contrast to the worldline particles they and their Poincar\' e transformations exist.

\section{Conclusions}

The spacetime Heisenberg Lie algebra excludes relativistic particles with a definite mass. 
This important result does not depend on this or that method of quantization. The lemma follows in each quantum theory by elementary algebra. The same conclusion
follows from the Stone-von Neumann theorem. 

More generally, each solution of a constraint which restricts an absolutely continuous spectrum to a lower dimensional submanifold
vanishes.


In quantum physics not only the algebra of operators is important but also the domain on which they act. 
As the spacetime Heisenberg Lie algebra does not allow a subspace of relativistic particles we specify the generators of Lorentz transformations
which do.

Though our lemma has far reaching implications its proof is astonishingly simple. That it had been overlooked for decades
by a multitude of researchers, authors, teachers and students is irrelevant for the correctness of the arguments, but noteworthy in
the history of science and for the sociology of scientific communities. 

\section*{Acknowledgements}
Norbert Dragon thanks Gleb Arutyunov, Arthur Hebecker and Hermann Nicolai for helpful email correspondence and  Wilfried Buchm\"uller,
Stefan Theisen and Sergei Kuzenko for extended, clarifying discussions.

\end{document}